\documentstyle[12pt,fleqn]{article}
\def\etal{{\it et al.\ }}                               
\def\mpc{{h^{-1}\,{\rm Mpc}\,}}

\begin{document}
\renewcommand{\thefootnote}{{\fnsymbol{footnote}}}
\setcounter{footnote}{1}

\pagestyle{myheadings}
\markright{$\xi_{cc}(r)$ for X--ray selected Abell clusters}

\noindent{\Large\bf
The Spatial Correlation Function From An X--ray} \\
\noindent{\Large\bf Selected Sample of Abell Clusters}

\vspace*{1.0cm}

\noindent{\large\bf R. C. Nichol$^1$,
U. G. Briel$^2$ \& J. P. Henry$^3$}

\vspace*{0.1cm}

\noindent$^1$Department of Physics \& Astronomy, Northwestern University, 2135
N.
Sheridan Road, Evanston, Il-60208, USA.$\footnote{Present address: Dept. of
Physics \& Astronomy, University of Chicago, 5640 S. Ellis, Chicago, Il-60637,
USA.}$ \\
\noindent$^2$Max--Planck--Institut f\"ur extraterrestrische Physik,
D--85740 Garching bei M\"unchen, Federal Republic of Germany.\\
\noindent$^3$Institute for Astronomy, 2680 Woodlawn Drive, Honolulu, Hawaii
96822,
USA.\\

\vspace*{0.05cm}

\begin{abstract}
We present here the Spatial Two--Point Correlation Function for
a complete sample of 67 X--ray selected Abell clusters of
galaxies. We find a correlation length of $16.1\pm3.4\mpc$ with
no significant clustering beyond $\simeq40\mpc$. This is the
lowest uncorrected value for the correlation length ever derived
from the Abell catalogue of clusters.  In addition, we have
investigated the anisotropy of the correlation function between
the radial and transverse directions. This can be characterised
by the magnitude of pair--wise cluster peculiar velocities such
anisotropy predicts, which we find to be $\simeq800{\rm
km\,s^{-1}}$ for our sample. Again, this is the lowest
uncorrected value ever seen for the Abell catalogue. We
therefore, no longer need to invoke high cluster peculiar
velocities or line of sight clustering to understand the
correlation function as derived from an Abell sample of
clusters.  Furthermore, our result is consistent with recently
published correlation functions computed from automated
selections of optical and X--ray clusters. Therefore, we are now
approaching a coherent picture for the form of the cluster
spatial correlation function which will be used to place
confident constraints on theories of galaxy formation.
\end{abstract}


\section{Introduction}
Clusters of galaxies are key tracers of the large--scale
structure in the universe, since their typical separation is
$\sim 10\mpc
\!\footnote{\rm Throughout this paper, we use $H_o=100\,{\rm
km\,s^{-1}\,Mpc^{-1}}$ and $q_o=\frac{1}{2}$.}$.  The most
popular statistic used to quantity the distribution of clusters
has been the Spatial Two--Point Correlation Function
($\xi_{cc}(r)$), whose observed shape and amplitude have been
the centre of much debate in the astronomical literature over
the last 10 years.

The bench--mark in this area of study has been the $\xi_{cc}(r)$
derived by Bahcall \& Soneira (BS83, 1983). For a sample of 104
$R\geq1$ Abell clusters (Abell 1958), they found that the
correlation function had the form; $\xi_{cc}(r)=(r/r_o)^{-1.8}$
with a correlation length of $r_o=25\mpc$ and a positive tail
out to $150\mpc$. One of the most severe consequences of this
result was it suggested that clusters of galaxies were over 15
times more clustered than galaxies, indicating that both
galaxies and clusters can not simultaneously be fair tracers of
the underlying mass distribution.  Furthermore, the form of the
BS83 result has been a strong constraint on models of galaxy
formation. For example, the popular theory of standard Cold Dark
Matter (CDM, Davis
\etal 1985) can not reproduce the high level of clustering seen in the BS83
correlation function, or its positive tail out to high
separations (White \etal 1987). More recently, Lilje (1990) has
also shown that the BS83 result can not be accounted for in a
Hot Dark Matter dominated universe.

Over the past few years, the original BS83 $\xi_{cc}(r)$ has
been confirmed, to some degree, with much larger redshift
surveys of Abell clusters. Table 1 highlights all published
correlation analyses carried out on the Abell catalogue and the
correlation functions derived from these studies.  The largest
surveys presented in this table, those of Huchra \etal (1990),
Postman \etal (1992) and Peacock \& West (1992), all find a
correlation length for $\xi_{cc}(r)$ of $r_o\simeq21\pm3\mpc$
with a positive tail out beyond $r>50\mpc$.  Again, all these
results are in conflict with CDM predictions for $\xi_{cc}(r)$.

\begin{table*}[ht]
{\small
\caption{(a) Published determinations of $\xi_{cc}(r)$. The table includes the
original reference, the number of
clusters used,  the published value of the correlation length ($r_o$),   the
number density of clusters and finally, the slope used for the analysis.}
\centering
\begin{tabular}{lcccc}
Survey & No. & $r_o$ ($h^{-1}$ Mpc) & $n_c$ ($10^{-5}\,h^{3} {\rm Mpc^{-3}}$)
&$\gamma$ \\
\multicolumn{5}{c}{Abell determinations of $\xi{cc}(r)$} \\
Bahcall \& Soneira 1983       & 104 & 25.0  & 0.6 & $1.8$
         \\
Ling {\it et al.} 1986        & 104& $21.9^{+7.1}_{-5.1}$ & 0.6 & $1.7\pm0.17$
         \\
Postman {\it et al.} 1986     &136& $20.0^{+4.6}_{-5.7}$& & $1.8$
         \\
Postman {\it et al.} 1986     &1207& $24.0^{+2.3}_{-3.0}$&& $1.8$
           \\
Postman {\it et al.} 1986     &370& $42.0^{+8.0}_{-9.1}$&& $1.8$
              \\
Huchra {\it et al.} 1990      &145& $20.3^{+4.8}_{-5.1}$ & $\sim 1.0$ & $1.8$
            \\
Huchra {\it et al.} 1990      & 92& $20.9^{+6.7}_{-6.9}$ & $\sim 0.6$ & $1.8$
            \\
West \& van der Bergh 1991    & 64& $22.1\pm6.8$ & & $1.7\pm0.5$
           \\
Postman {\it et al.}    1992  &351& $20.0^{+4.6}_{-4.0}$ & 1.2 & $2.5\pm0.2$
          \\
Postman {\it et al.}    1992  &156& $23.7^{+7.9}_{-9.0}$ & $\sim 1.0$ &
$1.8\pm0.2$         \\
Peacock \& West 1992          &195& $21.1\pm1.3        $ & 0.7       &
$2.0\pm0.2$         \\
Peacock \& West 1992          &232& $20.6\pm1.5        $ & 0.9        &
$1.5\pm0.2$         \\
\multicolumn{5}{c}{Projection Corrected Abell determinations of $\xi{cc}(r)$}
\\
Sutherland 1988               &533& $14.0^{+4.0}_{-3.0}$& 0.6 & $1.8$
             \\
Dekel {\it et al.} 1989       &102& $\sim 15.0$ & & $1.8$
                 \\
Sutherland \& Efstathiou 1991 &113& $9.0$ & $\sim 3.0$ & $1.8$
            \\
Sutherland \& Efstathiou 1991 &145& $14.0$ & 0.7 & $1.8$
              \\
Efstathiou {\it et al.} 1992   &298& $\sim 13$ & 1.4 & $\sim 2.0$ \\
\multicolumn{5}{c}{Non--Abell determinations of  $\xi{cc}(r)$} \\
Lahav {\it et al.} 1989       &53& $21.0$ & & $1.8$
   \\
Dalton {\it et al.}    1992   &220& $12.9\pm1.4$ & 2.4 & $1.9\pm0.3$
              \\
Dalton {\it et al.}    1992   &93& $14.4\pm4.0$ & 1.1 & $2.0$
          \\
Nichol {\it et al.}    1992   &79& $16.0\pm4.0$   & 1.0 & $2.1\pm0.3$
\\
Romer {\it et al.}     1993   &$129$& $13.7\pm2.3$      &     & $1.9\pm0.4$ \\
Romer {\it et al.}     1993   &$129$& $15.6\pm2.4$      &     & $1.4\pm0.4$ \\
\multicolumn{5}{c}{This Paper} \\
Nichol {\it et al.}    1993   &67& $16.1\pm3.4$         & 0.8& $1.9\pm0.3$ \\
\end{tabular}
}
\end{table*}

\setcounter{table}{0}

\begin{table*}[th]
{\small
\caption{(b) Published determinations of $\xi_{cc}(r)$.
This table details the cluster samples used by the authors in Table 1a.
$RC$ is Richness Class and $D$ is Distance Class as originally defined by Abell
(1958).}
\centering

\begin{tabular}{ll}
{Survey}                  & {Comments} \\
\multicolumn{2}{c}{Abell determinations of $\xi{cc}(r)$} \\
Bahcall \& Soneira 1983       & ${RC\ge 1}$, ${D\le4}$ Abell clusters. \\
Ling {\it et al.} 1986        & same data as above.      \\
Postman {\it et al.} 1986     &  Abell statistical sample $z\le 0.1$. \\
Postman {\it et al.} 1986     &  All Abell ${ RC\ge 1}$ clusters (80\%
estimated redshifts).\\
Postman {\it et al.} 1986     & All Abell ${RC\ge 2}$ clusters (75\% estimated
redshifts).\\
Huchra {\it et al.} 1990      & Deep Abell survey.\\
Huchra {\it et al.} 1990      & Deep Abell survey, ${ RC\ge 1}$.\\
West \& van der Bergh 1991    & cD Abell clusters.\\
Postman {\it et al.}    1992  &All Abell clusters $m_{10}\ge 16.5$.\\
Postman {\it et al.}    1992  & ${RC\ge 1}$ Abell clusters $m_{10}\ge 16.5$.\\
Peacock \& West 1992          & Volume limited sample of ${RC\ge 1}$ Abell
clusters.\\
Peacock \& West 1992          & Volume limited sample of ${RC\ge 0}$ Abell
clusters.\\
\multicolumn{2}{c}{Projection Corrected Abell determinations of $\xi{cc}(r)$}
\\
Sutherland 1988               & Abell catalogue + projection correction. \\
Dekel {\it et al.} 1989       & ${RC\ge 1}$, ${D\le4}$ Abell clusters +
projection correction. \\
Sutherland \& Efstathiou 1991 & Shectman (1985) clusters + projection
correction.\\
Sutherland \& Efstathiou 1991 & Huchra {\it et al.} (1990) survey + projection
correction.\\
Efstathiou {\it et al.} 1992   & Postman {\it et al.} (1992) survey +
projection correction.\\
\multicolumn{2}{c}{Non--Abell determinations of  $\xi{cc}(r)$} \\
Lahav {\it et al.} 1989       & EXOSAT X-ray clusters.\\
Dalton {\it et al.}    1992   & APM survey, $R\ge 20$.\\
Dalton {\it et al.}    1992   & APM survey, $R\ge 35$.\\
Nichol {\it et al.}    1992   & EM survey, $R\ge 22$. \\
Romer {\it et al.}     1993   & ROSAT All-Sky Survey. \\
Romer {\it et al.}     1993   & ROSAT All-Sky Survey. Fit $r<35\mpc$\\
\multicolumn{2}{c}{This Paper} \\
Nichol {\it et al.}    1993   & ROSAT detections of the
Huchra {\it et al.} (1992) sample. \\
\end{tabular}
}
\end{table*}

However, several authors (Sutherland 1988, Sutherland \&
Efstathiou 1991 \& Efstathiou \etal 1992) have claimed that the
high correlation length of $\xi_{cc}(r)$ derived from the Abell
catalogue, and its southern counterpart (Abell, Corwin \& Olowin
1989), is due to systematic biases introduced by the subjective
manner in which these catalogues were constructed. Sutherland
argues that the catalogues are plagued by projection effects,
which he defines as; {\it 'angular correlations that are not due
to genuine clustering in redshift space'.} These projection
effects would therefore, result in the correlation function
being artificially elongated in the redshift direction since
there would be an excess of cluster pairs angularly close on the
sky but with very different redshifts.  When Sutherland
corrected the BS83 result for these projection effects, he found
the correlation length of the $\xi_{cc}(r)$ decreased to
$14^{+4}_{-3}\mpc$ (Table 1); a result far less discrepant with
models of structure formation.

The elongation of $\xi_{cc}(r)$ in the redshift direction was
originally noted by BS83 themselves, however, they claimed the
effect was due to large cluster peculiar velocities ($\simeq
2000\,{\rm km\, s^{-1}}$, Bahcall, Soneira \& Burgett 1986)
which would have the effect of smoothing $\xi_{cc}(r)$ in the
line of sight direction. More recently, Jing, Plionis \&
Valdarnini (1992) have simulated the effect and claim the true
cause of these redshift elongations is real line of sight
clustering. Clearly, there is little consensus within the
astronomical literature over the true nature of these redshift
elongations.

This debate has increased dramatically over the past two years
with the publication of $\xi_{cc}(r)$ from new, fully automated
selections of clusters. For example, Nichol \etal (1992) have
presented $\xi_{cc}(r)$ for 79 rich clusters selected
objectively for the Edinburgh/Durham Cluster Catalogue (Lumsden
\etal 1992) which constitutes the Edinburgh/Milano cluster redshift survey
(Guzzo \etal 1992). Each of these clusters has an average of 10
galaxy redshift measurements, thus removing the problems of
phantom clusters and spurious cluster redshifts. For this unique
database, they found a correlation length of $16\pm4\mpc$ with
no positive tail beyond $\simeq 40\mpc$. Furthermore, they
showed that $\xi_{cc}(r)$ was isotropic with little indication
of extensive redshift elongations as seen in the BS83 result.
The best fit from this result for the magnitude of pair--wise
cluster peculiar velocities was $442^{+398}\,{\rm km\,s^{-1}}$,
which ruled out the value given by Bahcall \etal (1986) at the
$>3\sigma$ level.

Finally, in addition to the {\it Sutherland} effect, several
authors have argued that the true frequency of phantom clusters
is also a serious problem in the Abell catalogue.  This is the
chance alignment of groups/galaxies along the line of sight that
give the impression of a rich cluster, as seen in 2--D.
Simulations of this phenomenon have claimed that as many as
$50\%$ of clusters, seen in 2--D, are spurious (Lucey 1983,
Frenk \etal 1990). However, recent results from Briel \& Henry
(1993) have shown that over 80\% of Abell clusters are X--ray
emitters above a flux limit of $10^{-11}\, {\rm erg\,s^{-1}}$,
thus strongly suggesting that this is not as severe a problem as
predicted and therefore, has a much lower significance than the
proposed projection effects.

In this paper, we present $\xi_{cc}(r)$ derived from an X--ray
selected sample of Abell clusters. The data are the combination
of the deep Abell sample of Huchra \etal (1990) and the
corresponding X--ray detections as given by Briel \& Henry
(1993). The motivation behind this work was a hope that the
X--ray data would provide a more robust sample of Abell clusters
and minimise any projection effects that might be present. In
the next section, we discuss the exact sample of clusters used
in the analysis. In Section 3, we derive the correlation
function for this sample and in Section 4, compute the spatial
number density of our clusters.  We investigate the degree of
anisotropy for our sample in Section 5 and finally, end the
paper with a discussion of our result in the light of previous
correlation functions.

\section{Sample of Abell Clusters}

The results given in this paper are primarily based on the data
published by Huchra \etal (1990). They presented the redshift
measurements for all 145 Abell clusters in the region
$10^{h}\leq{\rm Right\,\,Ascension}\leq15^{h}$ and
$58^{\circ}\leq {\rm Declination}\leq78^{\circ}$, which is at
high galactic latitude and has an effective volume similar to
that used by BS83.  The median redshift of the sample is
$z=0.17$ with a maximum redshift of $z=0.35$ and an estimated
redshift completeness limit of $z=0.24$. Finally, only
$\simeq25\%$ of the cluster redshifts were derived from a single
galaxy redshift measurement thus reducing the probability of
assigning clusters a spurious redshift.

Recently, Briel \& Henry (1993) presented the X--ray details,
obtained from analysis of the ROSAT All--Sky survey, for the
clusters from the Huchra \etal deep sample. They detected 66 of
the 145 clusters at the $3\sigma$ level above the background and
for these clusters, they presented the X--ray flux and
luminosity of the cluster.  For the remaining clusters an upper
limit on the flux and luminosity were given in their paper.
However, all of these remaining clusters do have a measured flux
and luminosity, just at a lower significance level.  We refer
the reader to Briel \& Henry (1993) for a complete description
of the X--ray data used here.

We used all the measured luminosities and X--ray detection
levels to create a subsample from the Huchra \etal (1990)
clusters for our correlation analysis.  In total, we selected 67
clusters with the criteria of $z\leq 0.24$, an X--ray luminosity
of $\geq10^{43}\, {\rm erg\, s^{-1}}\,\,\footnote{\rm We
computed our luminosities using $H_o=100\,{\rm
km\,s^{-1}\,Mpc^{-1}}$. This is in contrast to the value used by
Briel \& Henry. However, we were able to reproduce their
luminosities when using the same value of $H_o$.}$ and a
detection significance of $\sigma\geq2$ above the background.
The redshift cut corresponded to the completeness limit of
clusters given by Huchra \etal which they derived from the
observed space density of the clusters as a function of
redshift. The two other criteria were derived empirically to
obtain a balance between selecting legitimate massive systems,
while still having enough clusters to carry out a confident
correlation analysis.  Hence, we selected the clusters for our
analysis from the Abell sample of Huchra \etal based only on the
X--ray characteristics of these clusters (Briel \& Henry 1993).
We did not introduce any optical selection criteria (other than
those already intrinsic to the Abell catalogue) since, at some
level, these are the basis for supposed projection effects {\it
i.e.} the optical richness of distant clusters being
artificially boosted because of the presence of a nearby rich
cluster.

\section{The Spatial Two--Point Correlation Function}

The Spatial Two--Point Correlation Function is usually estimated
by comparing the observed distribution of cluster pairs with
that obtained from a random catalogue of clusters distributed
within the same survey boundaries as the data.  The advantage of
this technique is that it minimises the problems of edge
effects, as well as allowing any selection biases in the data to
be incorporated into the random catalogue.  The function is
therefore, represented by;

\begin{equation}
\xi_{cc}(r)=\frac{2\,N_r}{N_d}\, \frac{n_{dd}}{n_{dr}} - 1,
\end{equation}

\noindent where $N_d$ and $N_r$ are the number of data and random clusters
respectively and $n_{dd}$ and $n_{dr}$ are the number of
data--data pairs and data--random pairs within the separation
interval $r\pm\Delta r/2$ ($\Delta r$ is the binsize). Comoving
separations ($r$) between the clusters were calculated using the
standard Friedmann cosmology. The reader is referred to Davis \&
Peebles (1983) and/or Peebles (1980) for a full discussion of
the merits of this estimator over other possible forms.  In all
estimations of $\xi_{cc}(r)$ discussed below, we used 3 random
catalogues ($N_r=100\,N_d$) which were then averaged together to
obtain a single $\xi_{cc}(r)$ (see Nichol 1992).  The redshift
distributions for these random catalogues were constrained to
have the same distribution as the data, after it had been
smoothed with a Gaussian of width $3000\, {\rm km\,s^{-1}}$.
This removed the need for detailed modelling of the redshift
selection function of the data and was the same procedure as
originally used by Huchra \etal and many other authors (Postman
\etal 1992, Nichol \etal 1992). The overall form of
$\xi_{cc}(r)$ was insensitive to the exact value of the
smoothing width.  The correlation function was derived using
logarithmic binning with a binsize of $\Delta {\rm log\, r}
=0.1$ and a maximum separation of $300\mpc$. Again, this was the
same procedure as used by Huchra \etal and therefore, allowed
for an exact comparison between these results. Finally,
bootstrap resampling errors on all correlation functions were
computed using the method described by Mo, Jing \& B\"orner
(1992).

The correlation function derived from our X--ray sample of 67
Abell clusters is shown in Figure 1. Included in this plot are
the best fit forms of $\xi_{cc}(r)$ as computed by Huchra \etal
(1990) and BS83. It is worth noting here, that we re--computed
$\xi_{cc}(r)$ for all 145 clusters in the deep Huchra \etal
sample and obtained exactly the same result as published by
them. We fitted our $\xi_{cc}(r)$ using the model
$\xi_{cc}(r)=(r/r_o)^{\gamma}$ and obtained a best fit for the
correlation length and slope of $r_o=16.1\pm3.4\mpc$ and
$\gamma=1.9\pm0.3$ for the range $5\leq r\leq35\mpc$
($\chi^2=2.1$ for 3 degrees of freedom). In addition, our
$\xi_{cc}(r)$ passed through zero at $\simeq 40\mpc$ with no
significant clustering beyond this.


\section{The Number Density of Clusters}

It is vital to accompany any discussion of the correlation
function with an accurate measure of the number density of
clusters being analysed. This is due to claims that the
correlation length of $\xi_{cc}(r)$ is a strong function of the
mean separation between the clusters in question (see Bahcall \&
West 1992).  The basis for such claims can be seen in Table 1
where richer clusters ( {\it i.e.} $RC\ge2$) have a larger
correlation length than other estimates of $\xi_{cc}(r)$.
Therefore, it is important to derive the number density of our
sample so that we can confidently compare our results with the
correlation estimates in the literature.

The number density, $n_c$, of any survey can be computed using
the following formulae;

\begin{equation}
n_{tot}=n_c\int\frac{dV}{dz}\,{\rm S}(z)\,{\rm E}(z)\,dz,
\end{equation}

\begin{equation}
\frac{dV}{dz} = 4\,d\Omega\left(\frac{c}{H_o}\right)^3\,\,\,\frac{
\left(z-\sqrt{z+1}+1\right)^2}{\left(1+z\right)^{\frac{7}{2}}},
\end{equation}

\noindent(Kolb \& Turner 1990, $q_o=\frac{1}{2}$)
where $n_{tot}$ is the total number of clusters observed, S$(z)$
is the selection function, E$(z)$ is the evolution of $n_c$ and
$d\Omega$ is the solid angle subtended by the survey region.

In Figure 2, we have plotted the product of $n_c\,{\rm
S}(z)\,{\rm E}(z)$ as a function of redshift for our sample of
67 clusters. This was computed by dividing the observed redshift
histogram of our 67 clusters (after it had been smoothed with a
Gaussian of width $3000\,{\rm km\,s^{-1}}$) by Equation 3.  The
error bars shown are $\sqrt{dN}$, where $dN$ is the number of
clusters in each redshift bin.  From this figure, our sample is
consistent with a constant number density out to a redshift of
$z=0.15$, which strongly suggests that neither the selection
function or cluster density evolution are significant. Beyond
this redshift, the observed number density of our sample drops
by a factor of three.  This would indicates that we are
approaching the limit of the Abell catalogue, which is
consistent with previous authors' estimations for the
completeness depth of the Abell catalogue (BS83, Scaramella
\etal 1991) and/or, we are seeing X--ray evolution in our deep
sample of X--ray clusters ({\it i.e.} Henry \etal 1992).
Therefore, from Figure 2, the underlying number density of our
X--ray selected sample is $\sim0.8\times10^{-5}\,h^3\,{\rm
Mpc^{-3}}$, which is consistent with the number densities of
other Abell samples as shown in Table 1. More specifically, our
$n_c$ agrees well with the mean number density quoted by Huchra
\etal (1990) and appears to be midway between the observed $n_c$
for $R\ge0$ and $R\ge1$ Abell clusters (Peacock \& West 1992).

As a further check, we compared the mean richness of our
individual clusters with that of other Abell samples.  This was
achieved using the individual cluster galaxy counts given by
Abell \etal (1989) for the northern clusters. We found our mean
richness to be $R_{mean}=75\pm31$ which is in good agreement
with the mean richness of the Abell $RC\ge1$ clusters;
$R_{mean}=74$ (Bahcall \& West 1992). The scatter on our mean
reflects the fact that our X--ray selection criteria have
selected clusters over a wide range of Abell richnesses which
can be seen in the poor correlation between X--ray luminosity
and galaxy richness (see Briel \& Henry 1993).  Furthermore,
this scatter is consistent with the findings of Lumsden \etal
(1992) who showed that the true external error on the quoted
richness of individual Abell clusters was $\simeq35$
irrespective of richness and distant class; this is a factor of
two greater than the error quoted by Abell. This agreement would
suggest that overall, we would expect to see similar clustering
characteristics as $RC\ge1$ Abell clusters.


\section{Anisotropy of $\xi_{cc}(r)$}

At the centre of the debate over the true form of $\xi_{cc}(r)$
are the observed elongations in the redshift direction for
functions derived from the Abell catalogue. As discussed in the
Introduction, Sutherland (1988) and others claim that these
elongations are due to projection effects, while Bahcall \etal
(1986) claim they are the result of large cluster peculiar
velocities. Therefore, we investigated our $\xi_{cc}(r)$ as both
a function of transverse separation ($r_p$) and radial
separation ($r_z$). Figure 3 shows the contours of our
$\xi_{cc}(r_z, r_p)$, as well as those for the whole Huchra
\etal sample (132 clusters) and the $RC\ge1$ Huchra \etal
clusters (92 clusters) both with $z\le 0.24$.



The best method of quantifying the amount of anisotropy seen in
these contour plots is via the value of cluster peculiar
velocities they predict.  This can be achieved using the
function,

\begin{eqnarray}
\xi_{cc}(r_z, r_p) & = &
\frac{r_o^{\gamma}}{\sqrt{2\pi}\,\sigma_v}\,\times\nonumber \\
                   &   & \int^{+\infty}_{-\infty}
(r_p^2 + (r_z - x)^2)^{-\gamma/2}\, e^{-x^2/2\sigma_v^2}\, dx,
\end{eqnarray}

\noindent where $\sigma_v$ is the value of the pair--wise cluster peculiar
velocities. This expression is the product of convolving the
power--law form of $\xi_{cc}(r)$ with a Gaussian dispersion.
Using the values of $r_o$ and $\gamma$ derived above, we fitted
our $\xi_{cc}(r_z, r_p)$ with the transverse direction
constrained to $0\leq r_p\leq 10\mpc$ and obtained a best fit
value of $\sigma_v=789^{+432}\, {\rm km\, s^{-1}}$ ($70\%$
confidence limit).  We also performed the same analysis on the
full sample of deep clusters and the $RC\ge1$ clusters, both
with $z\leq 0.24$, as shown in Figure 3. Using the Huchra \etal
values of $r_o$ and $\gamma$, we obtained
$\sigma_v=1241^{+342}\, {\rm km\, s^{-1}}$ and
$\sigma_v=1178^{+278}\, {\rm km\, s^{-1}}$ for the full and
$RC\ge1$ Huchra \etal datasets respectively.  These values are
higher than those quoted by Huchra \etal, but are still
substantially less than that advocated by Bahcall \etal (1986).
Due to the large statistical errors on these data, all the
results are within the 70\% confidence limits ($\simeq1\sigma$)
of each other.  It should be noted that all $\xi_{cc}(r_z, r_p)$
derived in this paper did have a local minima in the fitted
chi--squared at zero peculiar velocities, but these minima were
all larger than the eventual best fitted peculiar velocity
values.

\section{Discussion}

The value of $r_o$ computed from our sample of 67 X--ray
selected Abell clusters is the lowest uncorrected value of the
correlation length ever derived for the Abell catalogue.
Although other authors have derived lower values for $r_o$ from
the Abell catalogue (Table 1) this has been achieved by
correcting the correlation function for projection effects which
remains controversial and leads to large uncertainties on the
correlation length (Dekel \etal 1990).  In Figure 1, our
$\xi_{cc}(r)$ is systematically below both the Huchra \etal and
BS83 results on all scales.  If we constrain the comparison of
our result to samples with a similar number density ({\it i.e.}
$RC\ge1$), we still find we have a lower value of $r_o$ and a
lower degree of anisotropy. For example, we can compare our
anisotropy plots with that observed in similar samples of Abell
clusters without X--ray selection (Nichol \etal 1992, Peacock \&
West 1992 and the $RC\ge1$ Huchra
\etal sample presented in Figure 3c).  The $\xi_{cc}(r_p,
r_z)=1$ contour in these latter samples is clearly extended by
more than a factor of 2:1 in the redshift direction for $RC\ge1$
clusters. The same contour in our anisotropy plot is isotropic
with little sign of extension (Figure 3a). This difference is
quantified by the drop in the predicted cluster peculiar
velocities we see between our sample and the aforementioned
works.  Furthermore, our result is in good agreement with the
$\xi_{cc}(r)$ calculated by Nichol
\etal (1992) for an automated selection of galaxy clusters,
which has a similar number density to the sample used here, and
other non--Abell determinations (Table 1).

It should be stressed here, that the statistical errors on all
the measurements of $r_o$, either from Abell samples or not,
make it difficult to conclusively argue the difference seen in
Table 1 or in our result. No result is more than $2\sigma$ away
from any other determination of $r_o$. However, we feel that the
combination of a lower $r_o$ and a smaller degree of anisotropy
seen in our sample, suggests that this is a real change in the
clustering characteristics of our Abell sample. Moreover, this
debate should not be restricted to an exclusive discussion of
the correlation length; we also see different clustering
characteristics for our clusters on scales $>r_o$.

At face value, it is no longer necessary to invoke high cluster
peculiar velocities (Bahcall \etal 1986) or line of sight
clustering (Jing \etal 1992) to explain our result. If these
physical effects were real, it is hard to understand why we no
longer see them in our sample. Our lower value for the
anisotropy is probably due to the combination of two effects:
{\it (i)} we are selecting clusters based solely on their
X--rays luminosity which is much less prone to error than the
projected 2--D galaxy counts; {\it (ii)} a large majority of the
Huchra \etal clusters have multiple redshift measurements, thus
reducing the problems of assigning a cluster a spurious redshift
which would have the effect of smoothing the correlation
function preferentially in the redshift direction.

We can compare our result with other X--ray determinations of
$\xi_{cc}(r)$ in the astronomical literature (see Table 1).
Lahav \etal (1989) have published an $\xi_{cc}(r)$ for a
similarly sized sample of X--ray bright nearby clusters of
galaxies (53 clusters in total).  They found a correlation
length of $r_o=21\mpc$ which is in agreement with the standard
Abell value (Table 1) and therefore, appears to be in conflict
with other non--Abell samples of clusters and the findings of
this paper. However, the authors themselves comment that their
sample may be incomplete near the galactic plane and when they
curtail their sample to high galactic latitudes, they find an
$r_o$ of $17\mpc$ thus removing the apparent disagreement.
Furthermore, they force their correlation function to have a
predetermined slope of $\gamma=1.8$, which will have the effect
of increasing $r_o$ compared to that determined from a steeper
slope like $\gamma=2$ (${\rm log}\,r_o\propto \gamma^{-1}$).
More recently, Romer
\etal (1993) have presented the long awaited determination
of $\xi_{cc}(r)$ from a large objective sample of ROSAT
clusters. They find a correlation length of
$r_o=15.6\pm2.4\mpc$, over the same fitting range, with no
elongations in the redshift direction.  This is in good
agreement with the results presented here and results derived
from automated optical cluster catalogues (Table 1).

Bahcall \& West (1992) have argued for the existence of a
universial correlation function whose $r_o$ is directly related
to the number density of the clusters being analysed
($r_o=0.4\,n^{-\frac{1}{3}}$ where $n$ is the observed number
density).  Their arguement is analogous to the idea that richer
clusters have a higher correlation length, since they are
intrinsically rarer objects and thus have a larger mean
separation.  From the Bahcall \& West paper, the observed number
density presented in Section 4 for our sample of 67 clusters
would imply an observed $r_o$ of $20\mpc$ for this sample.  Our
result would therefore, appear to contradict this hypothesis
since we observe a lower correction length than that predicted.
However, the large error on both our observed $r_o$ and the
number density of our sample makes it hard to conclusively argue
this point.  As a matter of interest, the correlation length for
the remaining 78 clusters from the 145 Huchra \etal sample not
selected by our X--ray selection criteria is $10.8\pm4.4\mpc$
with a slope of $2.2\pm0.8$. This lower value of $r_o$ is
consistent with the idea that the correlation length is
dependent on the mass of the clusters analysed as we would
expect these 78 clusters to be, on a whole, less massive systems
than those in our main sample based on their X--ray luminosity.

Our result removes the tentative disagreement seen in Table 1
between Non--Abell and Abell samples of clusters, thus
suggesting that we are now approaching a coherent picture for
the distribution of clusters and their peculiar velocities.
Therefore, this allows us to use the correlation function of
clusters to confidently place tight constraints on theories of
galaxy formation. This has recently been carried out by several
authors ({\it i.e.} Olivier \etal 1993, Dalton \etal 1992) all
of which still find an excess of power on large scales compared
to predictions from the popular standard CDM model. However, the
discrepancy is far less severe than that seen with the original
BS83 result. In addition, there are now several alternative CDM
models available that might be able to reconcile this theory
with these new cluster correlation observations (see Bahcall \&
Cen 1992 \& Mann, Heavens \& Peacock 1993 and Croft \&
Efstathiou 1993).

\section{Acknowledgements}

Bob Nichol thanks Mel Ulmer for his support and advice during
the analysis of this data. We would also like to thank Chris
Collins and Luigi Guzzo for many simulating discussions and for
critically reading an earlier draft of this paper.  We are
grateful to an anonymous referee for his/her comments on this
paper.  Pat Henry and Ulrich Briel acknowledge support from NATO
grant CRG 910415 (JPH and UGB) and NASA grant NAG 5-1789 (JPH).

\end{document}